# Clinical Assistant for Remote Engagement Link (CARE-link): A Web-Based Electronic Health Records Software for Managing Diabetes

*Adjei Prince Ebenezer[a,b,c], Tettey Teye Joshua[a,b], Musah Toufiq[b], Agbeve Audrey[a,b], John Amuasi[a,b,d]

[a]Global One Health Research Group, Bernhard Nocht Institute of Tropical Medicine, Bernhard-Nocht-Str. 74, 20359 Hamburg, Germany; *emails: pe.adjei@kccr.de, joshua.tettey@kccr.de, audry.agbeve@kccr.de, amuasi@kccr.de*

[b]Global Health and Infectious Diseases Research Group, Kumasi Centre for Collaborative Research in Tropical Medicine, P.O. Box 1936, Kumasi, Ghana; *email: toufiqmush32@gmail.com*

[c]Department of Computer Engineering, Kwame Nkrumah University of Science and Technology, PMB, Kumasi, Ghana;

[d]Department of Global Health, School of Public Health, Kwame Nkrumah University of Science and Technology, PMB, Kumasi, Ghana

*Corresponding author: pe.adjei@kccr.de

## Abstract

CARE-link is an open-source, web-based clinical support platform designed to improve the management of gestational diabetes by linking clinicians and patients through an LLM-mediated workflow. The system aggregates patient-generated data outside the hospital, summarizes relevant clinical information, and delivers context-aware decision support to clinicians. For patients, CARE-link provides clear explanations of management plans and delivers timely lifestyle guidance through a WhatsApp interface. The integrated dual-facing design aims to promote continuous monitoring, support individualized care, and reduce the burden of in-clinic follow-ups. Built with a modular architecture, the platform can be adapted to other chronic conditions requiring longitudinal tracking and behavioral support. CARE-link has the potential to enhance clinical oversight, promote patient compliance, and strengthen continuity of care particularly in resource-constrained settings.

## Metadata

| Nr | Code metadata description | *Metadata* |
|---|---|---|
| C1 | Current code version | *V1.0* |
| C2 | Permanent link to code/repository used for this code version | *https://github.com/Digital-Health-and-AI-Team-AG-Amuasi/carelink.git* |

| C3 | Permanent link to reproducible capsule | *Not available* |
|---|---|---|
| C4 | Legal code license | *Creative Commons Attribution-NonCommercial (CC BY-NC) license* |
| C5 | Code versioning system used | *git* |
| C6 | Software code languages, tools and services used | *PHP, Laravel, Python, FastAPI, Langchain, ChatGPT API, WhatsApp Cloud API* |
| C7 | Compilation requirements, operating environments and dependencies | *Python 3.12, PHP 8.3, Laravel 12, FastAPI, Windows, Linux, macOS* |
| C8 | If available, link to developer documentation/manual | *https://docs.carelink.apomudenai.com/* |
| C9 | Support email for questions | *joshua.tettey@kccr.de* |

## 1. Motivation and significance

Effective management of chronic diseases (particularly diabetes - a major chronic condition that imposes strain on healthcare systems worldwide) relies on sustained patient engagement and lifestyle change [1]. In practice, many patients do not receive ongoing motivation or timely feedback on their behaviors between clinic visits, leading to preventable complications and poor outcomes. For managing chronic conditions like diabetes, Digital Behavior Change Interventions (DBCIs) that drive behavioral change outside traditional encounters, have been shown to be effective [2].

On the other hand, existing electronic health records (EHRs) typically fall short of providing context-rich and actionable insights at the point of care or to patients directly[3], [4]. Excessive documentation requirements leave providers with little time for patient dialogue, and EHR interfaces do not natively incorporate patient-generated data. Thus, EHRs generally offer one-way recording of medical data but lack built-in mechanisms to interpret those data in context or to send timely guidance back to patients[5]. This limitation means that clinically relevant changes often go unnoticed until the next visit, and opportunities for proactive intervention are missed.

In contrast, mobile instant-messaging platforms like WhatsApp have become widespread communication channels in low- and middle-income countries [6] with users spanning broad literacy and socioeconomic groups. Because it supports text, audio and multimedia messaging over affordable mobile networks, WhatsApp overcomes many traditional access barriers[6], [7] and has the potential to support behavioral change: it offers low-cost, real-time, two-way communication that can reinforce education, reminders, and self-monitoring prompts in the patient's own language and context.[8]

The Clinical Assistant for Remote Engagement link (CARE-Link) tool is motivated by the convergence of these needs and opportunities[9]. It employs a large language model (LLM) as an intelligent bridge between an EHR system and messaging platforms[10], [11], [12]. By continuously ingesting the patient's clinical record and real-time reports, CARE-Link's LLMs can generate personalized messages or feedback on demand. This approach follows recent demonstrations [13], [14] but extends it to a closed-loop format for chronic care. In summary, CARE-Link aims to overcome EHR and patient-engagement gaps by linking the widespread use and interactivity of mobile messaging with adaptive AI processing, delivering context-aware, actionable support to patients and clinicians in real time[13].

## 2. Software description

CARE-Link integrates an EHR with natural-language patient–clinician messaging, using LLMs to moderate a WhatsApp chat interface. Doctors interact via a web-based dashboard, while patients use WhatsApp. The system personalizes responses using each patient's medical context and delivers tailored education, reminders, and summaries relating to the clinician approved treatment plan. It supports continuity of care by using stored conversation logs, allowing context-aware decision support. The overall design relies on open-source technologies and modular services, facilitating extensibility and deployment in resource-constrained clinics. The overall implementation is summarized in Fig. 1.

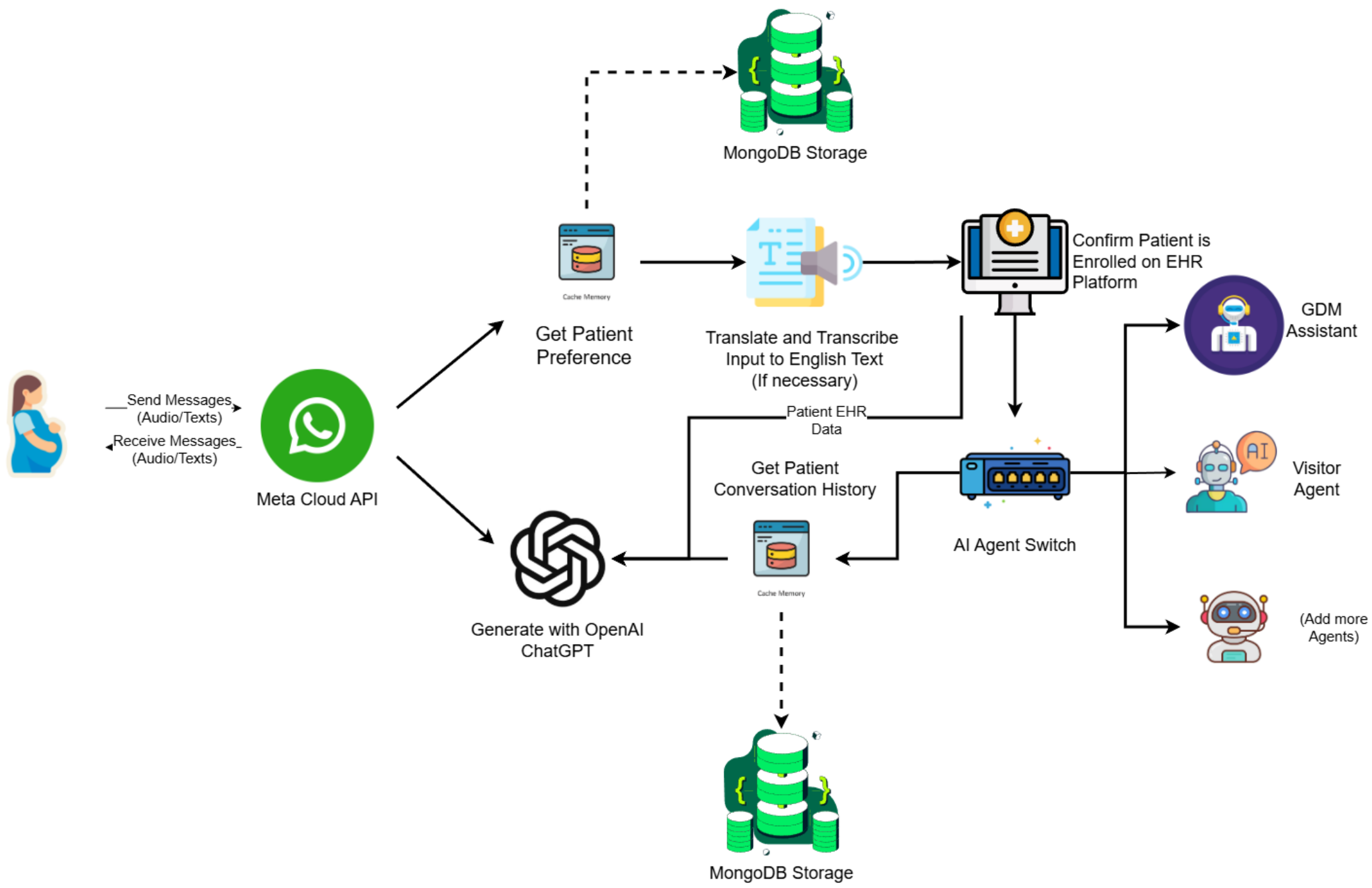

Figure 1: Illustrates the end-to-end data flow and microservice interactions underlying the CARE-Link platform, highlighting how patient-generated messages from WhatsApp are processed, analyzed, and integrated into the EHR.

### 2.1. Software architecture

CARE-Link is built on a service-oriented architecture that decouples the EHR, messaging, and AI components. Major components include:

**User interfaces:** A React-based web front-end for clinicians, and a WhatsApp chatbot interface for patients.

**EHR backend:** A Laravel (PHP) web service handles user management, patient records, and clinical visits. All patients and encounter data (demographics, vitals, complaints, interventions, etc.) are stored in a PostgreSQL database on Azure. RESTful APIs (with token-based authentication) mediate access between components.

**Chatbot service:** Python FastAPI application processes incoming WhatsApp messages. FastAPI's asynchronous, ASGI design allows handling many concurrent chat sessions.

Endpoints parse patient messages, manage dialog state, and post updates to the relevant database. The service also sends outgoing messages (educational content, reminders, summaries) via the Meta (Facebook) Cloud API for WhatsApp Business to patients.
**AI integration:** The Chatbot service (FastAPI application) queries an LLM (using the OpenAI ChatGPT API) to generate answers and suggestions. Domain-specific prompt templates incorporate patient data pulled from the database. For example, the clinician-facing assistant fetches relevant patient history or lab results and feeds these into LLM prompts so that generated advice is context-aware. Chat logs and patient data serve as "memory" to the LLM, enabling personalized responses over multiple sessions.

- **Deployment:** The Electronic Health Records (EHR) Platform and the Chatbot Service are deployed as decoupled, independently managed microservices. They communicate over secure HTTP/2 channels with end-to-end TLS encryption. The EHR Platform is implemented as an Azure App Service, while the Chatbot is hosted as an Azure Function for scalable, event-driven execution. Persistent data is maintained using Azure Cosmos DB for MongoDB workloads and Azure Database for PostgreSQL, providing high availability, low-latency access, and horizontal scalability.
- **Scalability and extensibility**: Each microservice can scale independently (for example, the FastAPI chat service can be load-balanced to handle many concurrent WhatsApp users). New agents can be added by deploying additional containerized services (e.g. a nutrition coach or lab-result explainer) that integrate via the same API framework. The use of standard interfaces (REST, JSON) and open-source stacks (React, Laravel, FastAPI, MongoDB, PostgreSQL) makes CARE-Link extensible and adaptable to other conditions or regions.

### 2.2. Software functionalities*:*

CARE-Link's core functionality spans across patient enrollment, clinical data management, AI-driven communication, and decision support. Key features include:

- **User management and authentication:** The system supports two user roles: clinicians and patients. Clinicians authenticate via secure web portal using passwords and add patients, while patients are identified by their registered phone numbers on WhatsApp.
- **Patient registration and medical records:** During the first antenatal visit, clinicians use the web interface to register a new patient. A structured intake form captures personal, demographic, and medical information. This information is stored in a structured form on PostgreSQL database. For each subsequent visit, clinicians fill out electronic forms recording the history of present illness, findings, investigations, impressions, and treatment plan. These visit records are also stored, enabling easy retrieval and analysis. Audit logs track all changes.
- **Patient-facing chatbot:** Enrolled patients interact with CARE-Link through WhatsApp. An LLM-powered assistant (prompts tailored to gestational diabetes) automatically engages with patients in a human-like manner. Its capabilities include education and Q&A, medication and visit reminders, emotional support and emergency detection.
- **Multimodal and local language support:** The patient-facing chatbot implements multimodal input handling, enabling both text-based and audio-based patient queries. The voice interface processes user-submitted speech prompts through an automatic speech recognition (ASR) pipeline, with current support for the Twi Ghanaian-language and English-language audio. This capability ensures robust interaction across varying literacy levels and communication preferences.

- **Clinician decision support**: Within the web EHR, clinicians have access to an AI query assistant. The system then composes a prompt combining the patient's structured data with the clinician's question and submits it to the LLM. The returned text is shown as a draft impression and recommendation (e.g. medication dose adjustment, lifestyle plan). The doctor can review and manually accept or modify these suggestions before applying them. This ensures full human oversight. Contextual knowledge (e.g. the patient's diet diary or repeated glucose logs) is incorporated so the advice is tailored. Prompt-based models have been shown to enhance medical decision support by contextualizing patient data.
- **Conversation analyzer agent:** During a patient visit, and as part of the clinical workflow, an AI agent is tasked with pre-filling the "Current Complaints" and "History of Presenting Complains" sections based on conversations between the patient and the GDM Assistant. A background AI agent continuously analyzes patient–bot chats for clinical signals. This semi-autonomous process ensures the care team stays informed of changes between visits. In effect, CARE-Link closes the loop: data flows from patient smartphone entries into the medical record, and from there into personalized clinical advice.
- **Security and privacy**: Every message exchange and data access occurs over encrypted channels (HTTPS/TLS). Role-based permissions and audit trails protect patient privacy. Patient phones see only anonymous content via WhatsApp, and no protected information is sent without encryption. All data resides in the hospital's cloud tenant; no patient data is shared with third parties beyond the LLM API (which transmits only prompt text, not raw identifiers). These measures align with HIPAA/GDPR principles.

## 3. Illustrative examples

This section presents representative use cases of CARE-Link within a typical clinical workflow, demonstrating how clinicians interact with the web-based EHR and how AI-generated insights assist data entry and decision-making.

### 3.1 Clinician Dashboard and Patient Management

After logging into CARE-Link, medical staff are directed to the Patient Management dashboard (Fig. 2). The page provides an overview of clinic activity, including the total number of registered patients, total visit count, and a "Priority Cases'' panel. This panel displays patients whose WhatsApp conversations have been flagged by the AI as potential emergencies or needing urgent follow-up. This allows clinicians to quickly identify high-risk individuals without manually checking each record. Fig. 3 presents the Priority Cases page, where clinicians can view detailed information about each flagged case.

The dashboard also supports streamlined navigation: clinicians can search for patients by name, phone number, or unique ID number, and can add new patients using a guided multi-step form. All patient records, including demographic details, obstetric history, vitals, exam findings, and past interventions, are stored and presented in structured form.

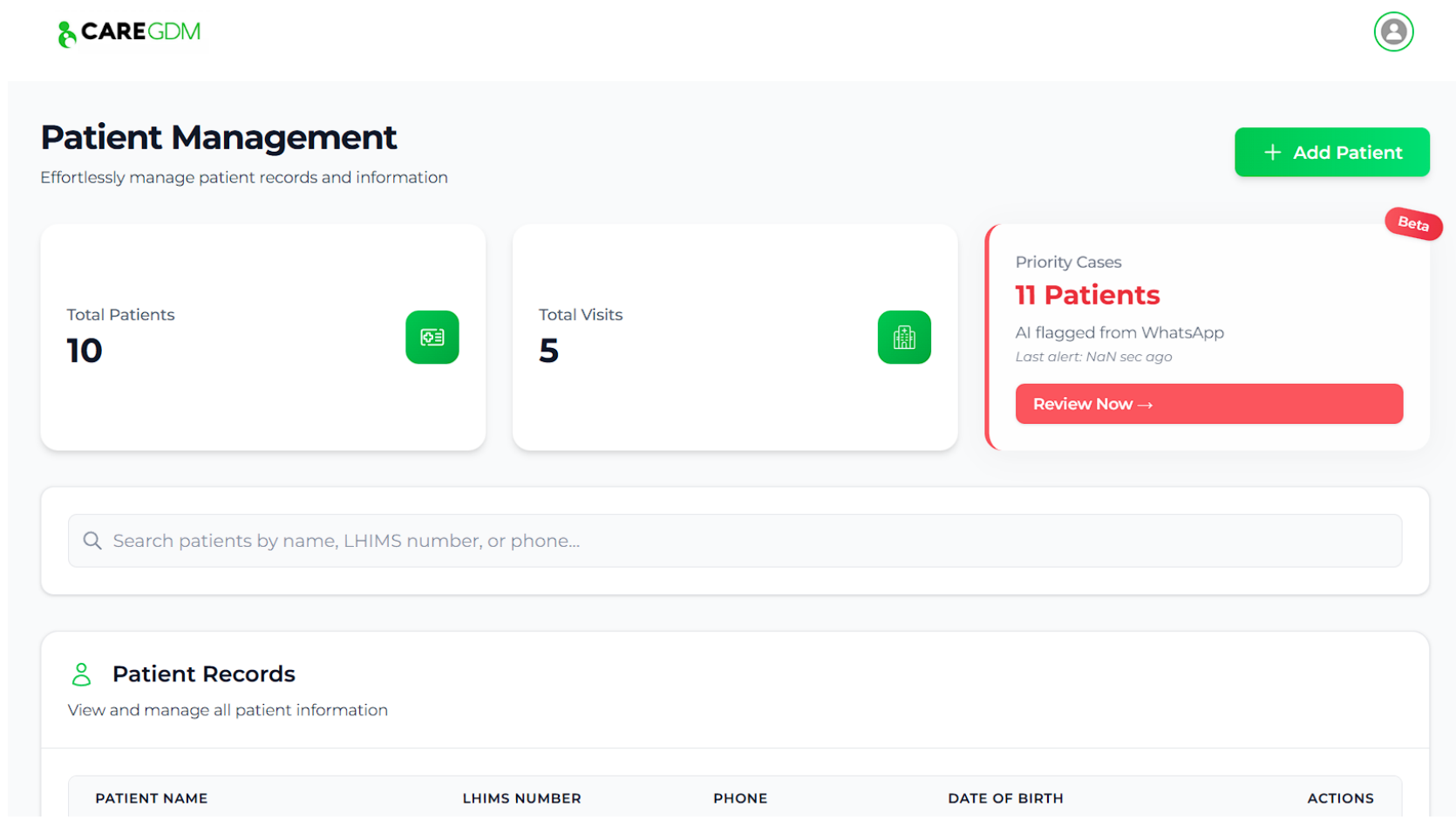

Figure 1: Patient Management Dashboard

### 3.2 Adding a New Clinical Visit

During a consultation, the clinician selects a patient and opens the Add New Record interface (Fig. 4). This form captures all components of a standard antenatal or diabetes follow-up visit. When the "Use AI" option is activated for relevant fields, the system automatically retrieves and processes the patient's recent WhatsApp conversations. A background conversation-analysis agent extracts medically relevant symptoms, behavioral patterns, and red-flag indicators, proposing draft summaries for the "current complaints" and "HPC" sections. The goal of this feature is to reduce clinician workload and ensure continuity between remote patient reports and in-clinic documentation.

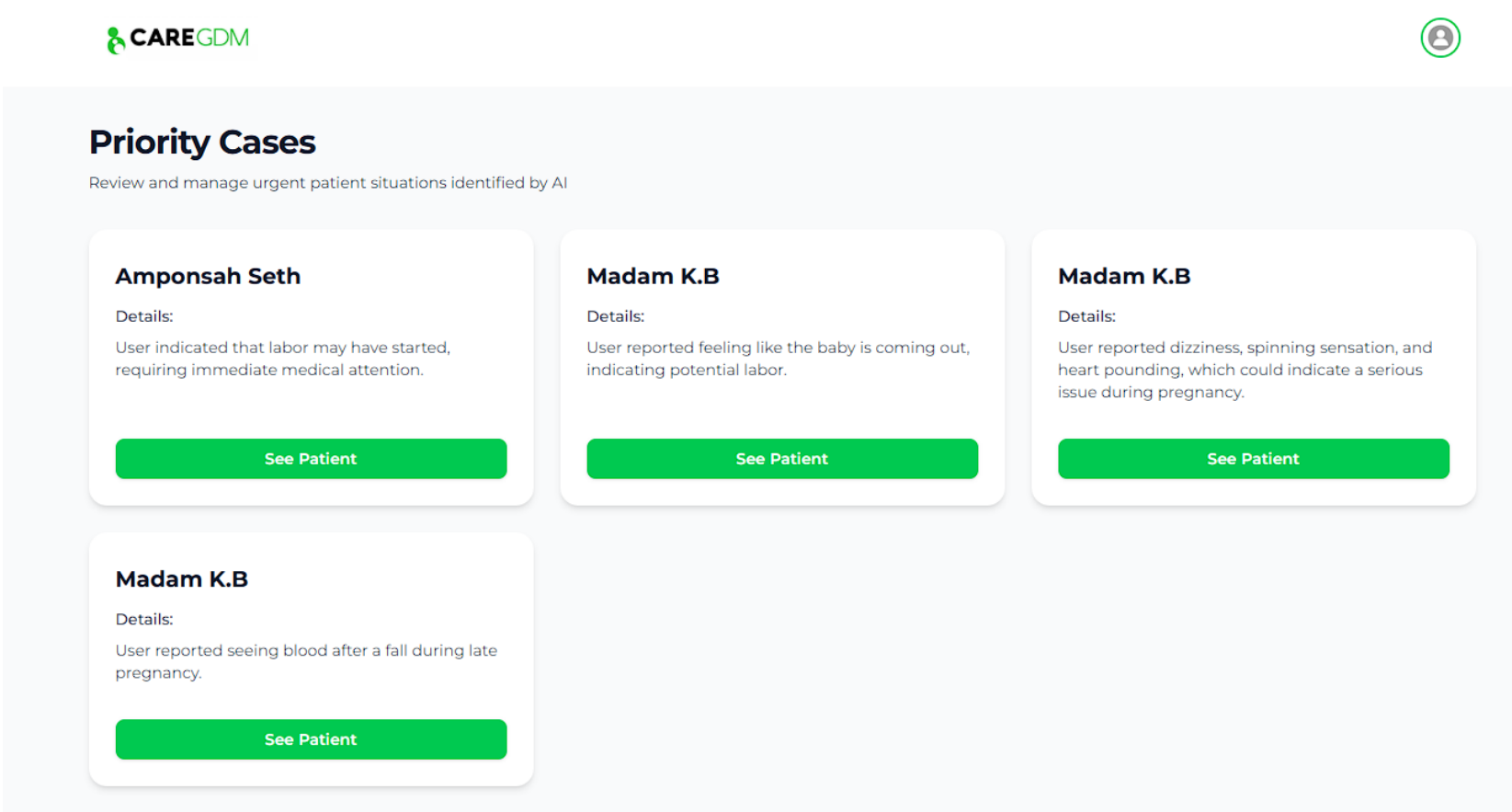

Figure 2: Page showing details of example emergency cases flagged by the agent

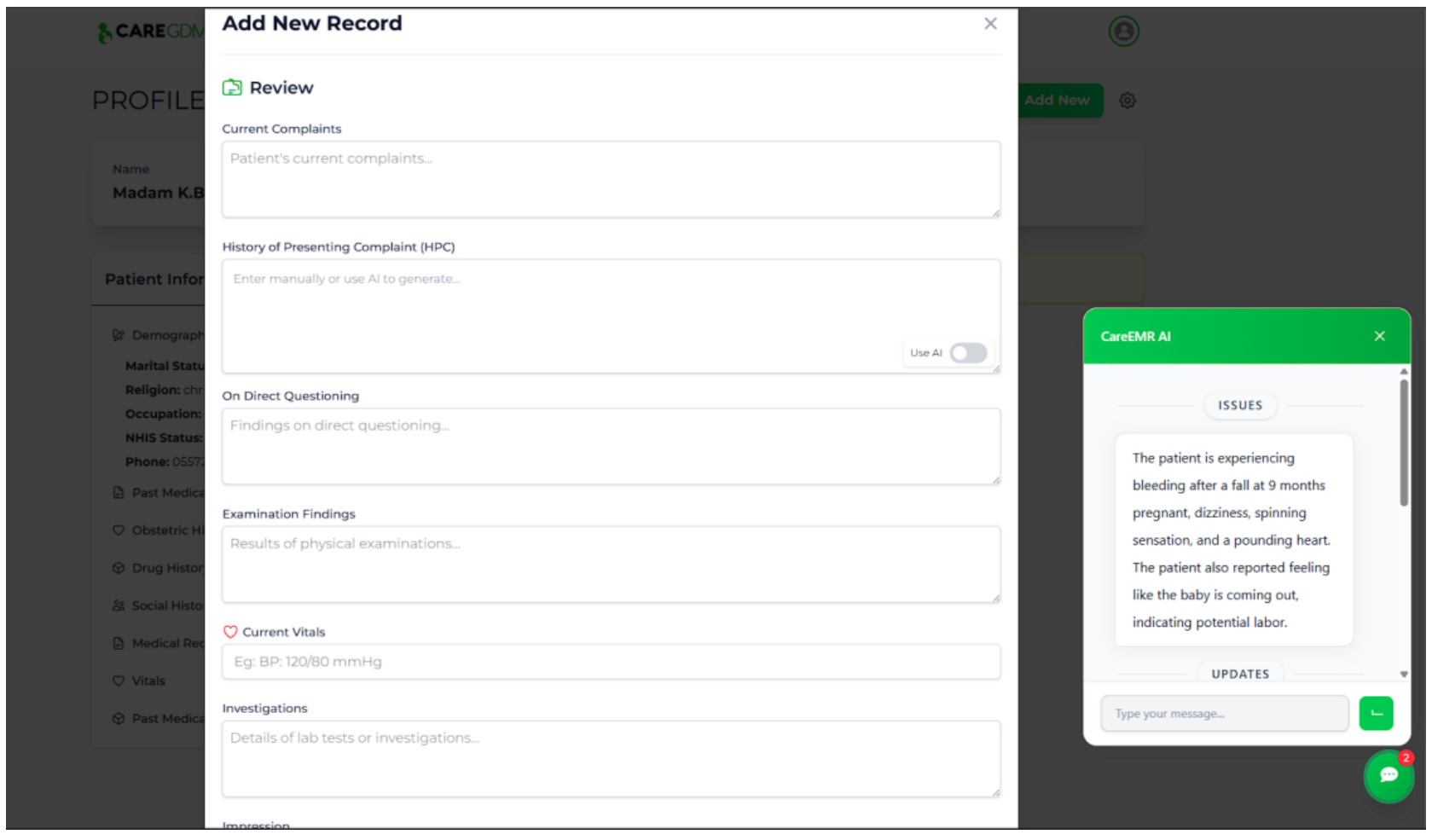

Figure 3: Add New Record Interface

### 3.3 AI-Assisted Review of Patient Conversations

A dedicated AI sidebar appears within the visit-entry interface (Fig. 5.c), summarizing patient-reported symptoms from recent WhatsApp exchanges. For example, in one case the assistant highlights that a patient (nine months pregnant) reported bleeding after a fall, dizziness, spinning sensation, and the feeling that "the baby is coming out." These symptoms are automatically flagged as potential emergencies and pushed to the clinician dashboard.

During the visit, the clinician can copy or refine these summaries and integrate them into the structured fields. This ensures that patient-generated information is accurately reflected in the medical record, while clinicians maintain full oversight of what is added.

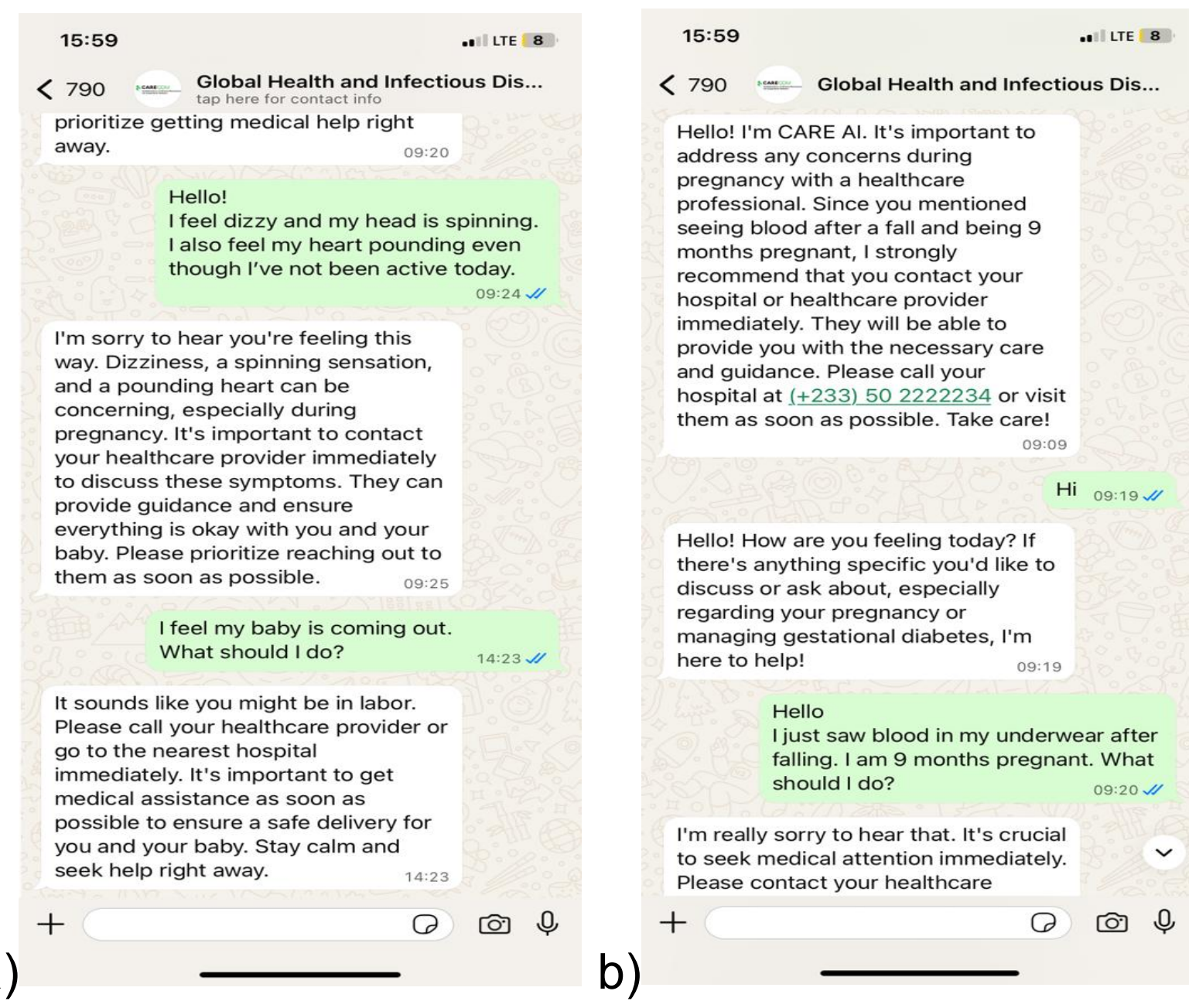

c) d)

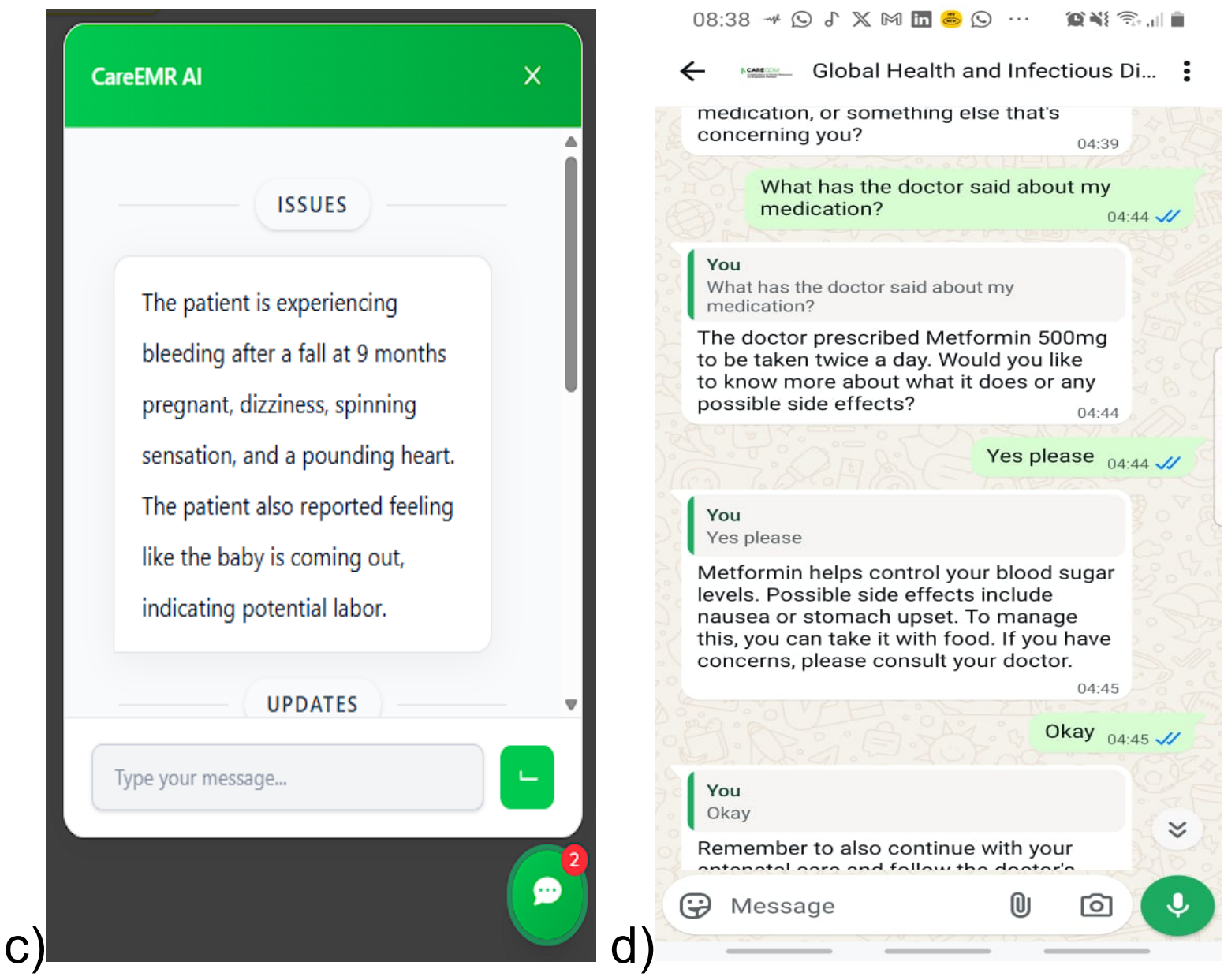

Figure 5: Panels (a) and (b) illustrate bidirectional message exchanges between the patient and the AI assistant, where patient-reported events are captured and context-specific guidance is generated in real time. Panel (c) depicts the clinician-facing AI summary sidebar, which aggregates and synthesizes patient-reported symptoms and contextual information extracted from the conversation stream. Panel (d) shows a subsequent patient inquiry seeking clarification on the clinician's recommended management plan.

### 3.4 Closing the Loop Between Chatbot and EHR

These examples illustrate CARE-Link's closed-loop workflow:

1. Patients communicate symptoms and daily updates via WhatsApp.
2. The AI agent analyzes conversations, identifies red flags, and pushes alerts to the EHR dashboard.
3. Clinicians review these findings while documenting a visit and may use AI-generated draft text to reduce manual entry.
4. Clinical decisions and updated treatment plans are stored, forming context for future AI-patient interactions.
5. System generates reminders to be sent out to patients at set times based on clinician decisions, recommendations or assigned medications.

This workflow demonstrates how CARE-Link seamlessly integrates patient-generated data into routine care, reduces documentation burden, and enhances clinical awareness between visits, particularly critical in managing gestational diabetes and other chronic conditions requiring continuous monitoring.

## 4. Impact

### 4.1 New Research Questions

CARE-Link opens critical research avenues at the intersection of AI, clinical communication, and health systems. First, how does LLM-mediated dialogue affect patient–clinician trust and therapeutic relationship? Emerging work suggests benefits for engagement, but potential risks to empathy and transparency[11]. Second, what design strategies best leverage WhatsApp's ubiquity to sustain patient adherence and engagement in LMICs[15], [16]. Third, how can generative AI in clinical workflows meet ethical obligations of beneficence, non-maleficence, and equity? Finally, what models ensure financial sustainability for AI-powered tools in resource-constrained systems[17]. CARE-Link provides a platform to empirically address these questions.

**4.2 Extension of Existing Research**

CARE-Link advances research across several domains. In patient-generated health data (PGHD), it extends prior app-based collection methods by integrating real-time WhatsApp messages and glucose logs into the EHR[18]. In clinical decision support (CDSS), it builds on LLM-based assistants that synthesize patient data and national guidelines to draft context-sensitive recommendations[19]. Unlike rule-based systems, CARE-Link leverages free-text interaction with clinicians and patients, enabling adaptive and human-like engagement. It also contributes to behavior change research by embedding AI-driven nudges (reminders, coaching) in a trusted messaging platform. This represents a next-gen fusion of PGHD integration, intelligent CDSS, and digital behavior change in LMIC settings.

**4.3 Commercial and Economic Impact**

CARE-Link proposes a dual-market strategy: free deployment in public systems and a "Premium Tier" for private hospitals offering expanded features. This aligns with value-based care principles: if AI tools improve outcomes and reduce complications, insurers or providers can recoup costs via savings [20]. CARE-Link also enhances workforce productivity by automating routine communication, enabling clinicians to focus on complex tasks. Comparable AI scribes have reduced workload and burnout in primary care [21]. Financing mechanisms, such as bundled payments or outcome-based reimbursements, could support adoption if real-world benefits are demonstrated [22]. CARE-Link is also well-positioned for public–private partnerships, following the path of other successful LMIC digital health ventures [23].

## 5. Conclusions

CARE-Link demonstrates an effective integration of EHR workflows, AI agents, and WhatsApp-based patient engagement. By combining structured clinical data with real-time patient reports, the system enhances monitoring, reduces documentation burden, and improves continuity of care. Its modular, cloud-ready architecture makes it adaptable to other chronic-care domains and scalable across resource-constrained settings.

### Acknowledgement

This work was funded by the Science for Africa Foundation, grant No. GCA/SFA/R15/053